\documentclass[12pt]{iopart}
\expandafter\let\csname equation*\endcsname\relax
\expandafter\let\csname endequation*\endcsname\relax
\usepackage{amssymb}
\usepackage{amsmath}
\usepackage{amsthm}
\usepackage[numbers,sort&compress,square]{natbib}
\usepackage{bm}
\usepackage{graphicx}
\usepackage{calc}
\usepackage[dvipsnames]{xcolor}
\usepackage{hyperref}
\usepackage{url}
\usepackage[capitalise]{cleveref}
\usepackage{tabularx}

\newcommand{\norm}[1]{\left\lVert#1\right\rVert}
\newcommand{\inner}[2]{\left\langle{#1,#2}\right\rangle}
\newcommand{\eexp}[1]{e^{#1}}
\begin{document}

\title[Dynamics in Bernoulli Space]{Classical and Quantum Dynamics in an Information Theoretic Space}

\author{Sean Golder$^{1,2}$ and Christopher Griffin$^{3,*}$}
\address{
	$^1$Department of Physics,
	The Pennsylvania State University,
    University Park, PA 16802
    }
\address{
	$^2$Department of Mathematics,
	The Pennsylvania State University,
    University Park, PA 16802
    }
\address{
	$^3$Applied Research Laboratory,
	The Pennsylvania State University,
    University Park, PA 16802
    }
\address{$^*$Corresponding Author}

\eads{\mailto{smg7235@psu.edu}, \mailto{griffinch@psu.edu}}
\date{\today~-~Preprint}

\begin{abstract} We study elementary classical and quantum dynamics in an information geometric space corresponding to a Bernoulli random variable, extending work by Goehle and Griffin [\textit{Chaos, Solitons \& Fractals}, 188, 115535, (2024)], who study the information theoretic analogue of the spring-mass system. Information geometric constructions are useful in both statistical physics and {geometric quantum mechanics}, and they play a role in areas of biophysics. In this letter, we derive the spectrum for the Laplace-Beltrami operator in Bernoulli space and find solutions to the wave {and} heat equations. We then show {one way} to quantize momentum in Bernoulli space {and identify appropriate boundary conditions and domains that ensure self-adjointness.} {We then} obtain energies and wavefunctions for {a} free particle. {Unlike the Euclidean space, Bernoulli space has at least two natural square distances, derived from the geodesic distance and the information theoretic Kullback-Leibler divergence.} {We} show that quadratic approximation of the {corresponding} Kullback-Leibler potential used by Goehle and Griffin results in a quantum oscillator in information space that is equivalent to a quantum pendulum in Euclidean space. {We use this to construct lower bounds on the energy levels of both the Kullback-Leibler and geodesic potentials.}
\end{abstract}

\vspace{2pc}
\noindent{\it Keywords}: Fisher metric, Bernoulli variable, quantum harmonic oscillator, quantum pendulum

\submitto{\jpa}
\maketitle

\section{Introduction} Nonlinear oscillators have been studied in both the classical and quantum regime by various authors. The early work of Matthews and Lakshmanan \cite{mathews1974unique} led to a nonlinear oscillator having both closed form solutions and yielding a  natural Hamiltonian structure, which was both extended to higher dimensions by Cari\~{n}ena \cite{carinena2004non,carinena2007quantum} and readily quantized \cite{carinena2004one}. Likewise, Calogero \& Graffi \cite{calogero2003quantization} and Calogero \cite{calogero2004quantization} quantize non-linear {harmonic} oscillators with Hamiltonian structure and natural closed form solutions, while Gubbiotti and Nucci study the quantization of a Li\'{e}nard-type nonlinear oscillator \cite{gubbiotti2014noether}, and Ghosh et al. quantize an isochronous oscillator \cite{Ghosh2025isochronous}. While most work is on undamped systems, Dekker \& Blacker \cite{dekker1977quantization} and Tilbrook \cite{blacker2021alternative} consider the quantization problem for nonlinear damped oscillators. In fullest generality, this body of work seems to fall into the class of nonlinear analogues of quantum field theory as studied by Delbourgo et al. \cite{delbourgo1969infinities} and Nishijima and Watanabe \cite{nishijima1972green}. Though more recent work considers quantization of non-Hamiltonian systems as well (see Chia et al. \cite{chia2025quantization}).

A rich class of nonlinear oscillators can be generated by simply considering a (non-Euclidean) Riemannian metric $g_{ij}$ along with a convex potential function $V(q)$ and using the Lagrangian,
\begin{equation}
    \mathcal{L} = \frac{1}{2}g_{jk}\dot{q}^j\dot{q}^k - V(q),
    \label{eqn:MainLagrangian}
\end{equation}
giving Newton's laws on the Riemannian manifold \cite{landau1976mechanics} corresponding to $g_{ij}$. For an appropriate choice of $g$ and $V$, this yields two benefits: (i) the force $F = -\partial_q V$ is a natural generalization of Hooke's law to the manifold and thus is simply the {harmonic} oscillator in a non-Euclidean geometry, and (ii) when the underlying manifold of $g$ exhibits curvature, studying the quantization of the corresponding Hamiltonian may provide insights into quantum mechanics in curved geometries, if not (highly) curved spacetimes.

In this letter, we apply the espoused principle by considering a metric tensor $g$ arising as the information geometric Fisher metric \cite{nielsen2020elementary} for the Bernoulli distribution. While the approach is fully generalizable, we select the Bernoulli distribution because it yields several simple one-dimensional analogues to the classical quantum {harmonic} oscillator. Interestingly, information spaces, like the Bernoulli space we consider, admit both (geodesic) square distances and consistent (Bregman) divergences \cite{amari2016information,nielsen2020elementary,nielsen2018information} that can be used as pseudo-square distances because of their relation to information projection \cite{nielsen2018information} in these spaces. Goehle and Griffin \cite{goehle2024dynamics} study an analogy of Hooke's law in Bernoulli space using \cref{eqn:MainLagrangian} in which the square distance from the metric is replaced by the Kullback-Leibler divergence. 

This letter is motivated by the following observations: {(i)} Crooks \cite{crooks2007measuring} has shown a natural relationship between thermodynamics and information geometry, suggesting the potential for a relationship between quantizations of the dynamics on information geometric manifolds and quantum thermodynamics. {(ii)} There has been recent work on the use of information geometric constructions in expressing aspects of Friston's free energy principle \cite{FKH06,F12,K12,FDSH23}, a {biophysical} formulation of the Bayesian brain hypothesis. {(iii) Many of the} Riemannian manifolds arising from information geometry have natural curvature, with regions of high curvature corresponding to regions of low variance, thus providing a natural way to study quantization in curved spaces. Before proceeding, it is worth noting that this work is related to, but quite distinct from work on quantum geometry \cite{braunstein1994statistical}, the quantum Fisher matrix \cite{vsafranek2018simple}, quantum statistics \cite{barndorff2000fisher}, and the relationship between classical probability, quantum mechanics, and the manifold of pure states \cite{facchi2010classical}. These works focus on geometric foundations of the unique probability structures found in quantum mechanics, while this letter is specifically interested in treating a Fisher manifold as a space on which to study classical and quantum dynamics.

The remainder of this letter is organized as follows. We provide necessary background on information geometric constructions in \cref{sec:Background}. In \cref{sec:Classical} we construct the Laplace-Beltrami operator {and find its corresponding eigenvalues and eigenfunctions assuming various kinds of boundary conditions.} We then use {these results} along with the methods from Cari\~{n}ena to construct a natural quantization of the {momentum and} Hamiltonian {with appropriate operator domains} and identify energy levels and eigenfunction solutions for the free particle problem in \cref{sec:FreeParticle}. The energy levels and eigenfunctions for a variant of the quantum {harmonic} oscillator in this geometry are derived in \cref{sec:HarmonicOscillator}. {We also use these results to characterize the behaviour of other oscillators in this space.} We provide conclusions and future directions in \cref{sec:Conclusion}.

\section{Background}\label{sec:Background}
Let $p(x;\eta)$ be a probability distribution parameterized by $\eta$. The Fisher metric arising from $p(x;\eta)$ is given by,
\begin{equation*}
    g_{jk} = \int_X dx \, \frac{\partial \log(p)}{\partial \eta^j}\frac{\partial \log(p)}{\partial \eta^k}p(x;\eta).
\end{equation*}
Let $\eta(\lambda)$ with $\lambda \in [\lambda_0,\lambda_f]$ be a path in distribution parameter (information) space. Following Crooks \cite{crooks2007measuring}, Nielson \cite{nielsen2020elementary}, or Amari \cite{amari2016information}, the kinetic energy action,
\begin{equation}
    A = \frac{1}{2}\int_{\lambda_0}^{\lambda_f} d\lambda \,g_{jk}\dot{\eta}^j\dot{\eta}^k,
    \label{eqn:KEAction}
\end{equation}
yields {an} entropy per unit ``time'' needed to traverse the path. Note the use of the Einstein summation convention. Subsequently, we will replace $\lambda$ with $t$ (time) and consider classical and quantum dynamics of a particle in this space. Among other unusual properties, each information manifold admits a (Bregman) divergence \cite{amari2016information} that is consistent with the metric. The Kullback-Leibler (KL) divergence is the natural such divergence in the exponential family of distributions, which includes the categorical distributions. Let $\eta$ and $\zeta$ be two vectors of distribution parameters. Following Goehle's notation \cite{goehle2024bayesian,goehle2025approximation}, the Kullback-Leibler divergence for two distributions in the same family is,
\begin{equation*}
    D_{KL}(\eta|\zeta) = \int_X dx\,p(x;\eta)\log\left[\frac{p(x;\eta)}{p(x;\zeta)}\right].
\end{equation*}
While this definition is usually (and readily) extended to arbitrary pairs of distributions, not necessarily in the same family \cite{amari2016information}, we will not need this generalization. The classic interpretation of the KL divergence is that it returns the extra information (in nats) needed to decode information using $p(x;\eta)$ when the data is encoded using $p(x;\zeta)$. Nielson \cite{nielsen2020elementary} provides a more geometric interpretation of the KL divergence. If $\mathcal{M}$ is the Riemannian manifold corresponding to $g_{jk}$ and $\mathcal{R}$ is a submanifold, then the KL divergence characterizes the projection from a point $\eta$ in $\mathcal{M}$ to the manifold $\mathcal{R}$, where $\zeta \in \mathcal{R}$ is the point minimizing the divergence. While KL divergence is not symmetric, preventing it from being a true metric, it acts as a natural projection square distance with,
\begin{equation*}
    D_{KL}(\eta|\zeta) \sim \norm{\eta - \zeta}^2 + O\left[\norm{\eta}^3\right] + O\left[\norm{\zeta}^3\right].
\end{equation*}
The KL divergence is locally consistent with the Fisher metric \cite{nielsen2018information} with the approximation,
\begin{equation*}
D_{KL}\left(\eta\vert\eta_0\right) =\frac{1}{2} \mathbf{g}_{jk}(\eta_0)\Delta\eta^j \Delta\eta^k + O(\norm{\Delta\eta}^3),
\label{eqn:fisherkl}
\end{equation*}
holding for small distances $\Delta\eta = \eta - \eta_0$. Thus, both the KL divergence and the Fisher metric measure square distance in nats (or bits).

For the Bernoulli distribution with (one-dimensional) parameter $q$, the Fisher information metric takes the simple one-dimensional form,
\begin{equation*}
    g_{ij} =g=\frac{1}{q(1-q)}.
\end{equation*}
{Note that the Riemann curvature tensor resulting from this metric will be zero. This is expected, as all one-dimensional manifolds lack curvature.}
%

To study classical and quantum dynamics in Bernoulli space, we first formulate the Laplace-Beltrami operator for this space as,
\begin{equation*}
    \Delta u = \frac{1}{\sqrt{|g|}}\partial _{i}\left[\sqrt{|g|}\ g^{ij}\partial_{j}u\right] = \sqrt{q(1-q)}\frac{\partial}{\partial q}\left[\sqrt{q(1-q)}\frac{\partial u}{\partial q}\right].
\end{equation*}

\section{Spectrum of the Laplace-Beltrami Operator}\label{sec:Classical}

Consider the Helmholtz equation,
\begin{equation*}
    \Delta u = -\omega^2 u,
\end{equation*}
in the Bernoulli geometry. Expanded, the Helmholtz equation is the (ordinary) differential equation,
\begin{equation*}
    \Delta u = \sqrt{q(1-q)}\frac{d}{dq}\left[\sqrt{q(1-q)}\frac{du}{dq}\right] = -\omega^2u,
\end{equation*}
which has Sturm-Liouville (SL) form with weight function,
\begin{equation*}
    w = \frac{1}{\sqrt{q(1-q)}}.
\end{equation*}

\begin{figure}[htbp]
    \centering
    \includegraphics[width=0.6\textwidth]{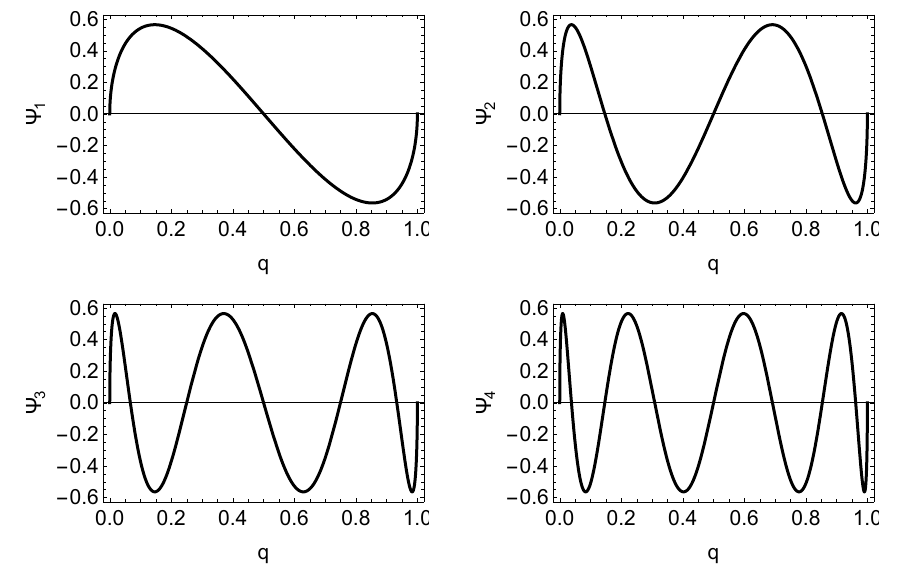}
    \caption{{We illustrate the imaginary part of the eigenfunctions for $n=1,2,3,4$.}}
    \label{fig:Eigenfunctions}
\end{figure}

{Integration on the manifold with respect to the Riemannian volume form can be thought of as integration on $[0,1]$ with respect to the $w$-weighted Lebesgue measure $\mu$, since,
\begin{equation*}
    dV_g=\sqrt{|g|}dq=\frac{dq}{\sqrt{q(1-q)}}=w(q)dq=d\mu(q).
\end{equation*}}
{Since $w$ is not both continous and bounded on the closed interval $[0,1]$, we cannot apply SL theory directly and instead must appeal to a more complete Hilbert space analysis, which we do in the next section.} {Despite this}, we can find a closed form solution to the Helmholtz equation using the (standard) projection of the square roots of the categorical distribution parameters to the positive orthant of the unit sphere \cite{gromov2014search} with,
\begin{equation}
    q = \sin^2\left(\frac{\theta}{2}\right) \qquad {\frac{d}{d\theta} = \sin\left(\frac{\theta}{2}\right)\cos\left(\frac{\theta}{2}\right)\frac{d}{dq}}.
    \label{eqn:Transform}
\end{equation}
Here, {$\theta\in [0,\pi]$}. Making this substitution transforms the Helmholtz equation in Bernoulli space to the ordinary Helmholtz equation in one dimension,
\begin{equation*}
    \frac{d^2u}{d\theta^2} = -\omega^2u.
\end{equation*}
{If we impose periodic} boundary conditions, {$u(0) = u(\pi)$ and $u'(0) = u'(\pi)$, the} eigenfunctions are the well-known Fourier basis, 
\begin{equation*}
    {u_n(\theta) = A_n \exp(2ni\theta),}
\end{equation*}
where {for} $n \in \mathbb{Z}$ {we have} the discrete spectrum of the operator, {$(2n)^2$}. Inverting the transformation yields the eigenfunctions of $\Delta$ in  Bernoulli space,
\begin{equation*}
    {u_n(q) = A_n\exp\left[4ni\arcsin(\sqrt{q})\right],}
    %
\end{equation*}
{which obey the corresponding periodic boundary conditions.} By construction these functions are orthogonal, i.e., $\inner{u_n}{u_m} = 0$, if $n \neq m$, which can easily be confirmed by direct computation. The normalization coefficient can be found from the fact that,
\begin{equation*}
{\inner{u_n}{u_n} = \int_0^1 dq\,\frac{\overline{u_n(q)}u_n(q)}{\sqrt{q(1-q)}} = \pi }
\end{equation*}
thus giving the orthonormal eigenbasis of the Bernoulli space Laplace-Beltrami operator,
\begin{equation}
    {\Psi_n(q) = \frac{1}{\sqrt{\pi}}\exp\left[4ni\arcsin(\sqrt{q})\right]},
    \label{eqn: period_basis}
\end{equation}
with $n \in \mathbb{Z}$. The {imaginary parts of the } first few eigenfunctions are shown in \cref{fig:Eigenfunctions}.

{These eigenvalues and eigenfunctions are as much a product of the periodic boundary conditions as the differential equation. Replacing these with homogeneous Dirichlet boundary conditions gives the (expected) eigenbasis,
\begin{equation}
    \Phi_n(q) = \sqrt{\frac{2}{\pi}}\sin\left[2n\arcsin(\sqrt{q})\right],
    \label{eqn: dirichlet_basis}
\end{equation}
with $n\in \mathbb{Z}$ and $n > 0$, and corresponding eigenvalues $n^2$. A similar analysis provides an eigenbasis with homogeneous Neumann boundary conditions.}  From this analysis, solutions to the heat and wave equations with homogeneous Dirichlet boundary conditions in Bernoulli space follow immediately as,
\begin{equation*}
    {\mathcal{K}(q,t) = \sum_{n=1}^\infty A_n\eexp{-nt}\Phi_n(q) \qquad \mathcal{W}(q,t) =\sum_{n=1}^\infty A_n\eexp{-int}\Phi_n(q),}
\end{equation*}
where $A_n$ are determined from initial conditions and the standard procedure for recovering generalized Fourier coefficients \cite{strauss2007partial} assuming initial conditions are given. {We now make use of the identified eigenvalues and eigenfunctions in the periodic boundary condition case to analyze simple quantum mechanical systems in Bernoulli space. As we show in the following section, the periodic boundary conditions will ensure both the energy and momentum operators are self-adjoint under appropriate domain assumptions.
}

\section{Quantum Mechanical Free Particle}\label{sec:FreeParticle}
Following Goehle and Griffin \cite{goehle2024dynamics}, we use a modified kinetic energy Lagrangian, 
\begin{equation*}
    \mathcal{L} = \frac{m}{2}g_{jk}\dot{q}^j\dot{q}^k=\frac{m}{2}\frac{\dot q^2}{q(1-q)},
\end{equation*}
to derive the equations of motion for a free particle, where the conjugate momentum is,
\begin{equation*}
    {p = \frac{\partial\mathcal{L}}{\partial \dot q}=\frac{m\dot{q}}{q(1-q)}.}
\end{equation*}
The Hamiltonian follows from the Legendre transform as,
\begin{equation*}
    \mathcal{H} = \frac{1}{2m}q(1-q)p^2.
\end{equation*}

While there are several approaches to quantizing this Hamiltonian, it is simplest to choose a quantization so that the kinetic energy operator retains the form,
\begin{equation*}
    \hat T = \frac{1}{2m}\hat p^2,
\end{equation*}
and remains analogous in structure to its classical counterpoint. Following the approach by Cari\~{n}ena et al. \cite{carinena2004one} we quantize the momentum as,
\begin{equation*}
    \hat p = -i\hbar \sqrt{q(1-q)}\frac{\partial}{\partial q}.
\end{equation*}
We then obtain the familiar expressions for $\hat p^2$ and for $\hat T$,
\begin{equation*}
    \hat p^2 = -\hbar^2 \Delta, \quad\text{and}\quad \hat T= -\frac{\hbar^2}{2m}\Delta,
\end{equation*}
where $\Delta$ is the Laplace-Beltrami operator in Bernoulli space. {For the position, $q$, we may simply use the standard quantization $\hat q \rightarrow q$.}

{
We now investigate the domain of the proposed momentum operator under the periodic boundary condition $\psi(0) = \psi(1)$. For continuously differentiable functions $\psi$ and $\phi$ on $[0,1]$, we have,
\begin{align*}
\langle \phi,\hat p\psi\rangle&=\int_0^1\overline\phi(\hat p\psi)\sqrt{|g|}dq=-i\hbar\int_0^1\overline{\phi(q)}\frac{\partial \psi}{\partial q}\sqrt{q(1-q)}\frac{dq}{\sqrt{q(1-q)}}\\
&=-i\hbar \left.\overline{\phi(q)}\psi(q)\right|_0^1+\int_0^1\overline{\left(-i\hbar\frac{\partial\phi}{\partial q}\right)}
\psi(q)dq=-i\hbar \left.\overline{\phi(q)}\psi(q)\right|_0^1+\langle \hat p\phi,\psi\rangle.
\end{align*}
Since the momentum operator should be symmetric, this implies \textit{a fortiori} that any function $\psi$ in the domain of $\hat p$ must satisfy $\psi(0)=\psi(1)$, justifying the imposed boundary condition. Thus, it is natural to posit that the domain of $\hat p$ is,
\begin{equation}
\operatorname{Dom}(\hat p)=\left\{\psi\in L^2([0,1],d\mu(q))\mid \psi \in C^1([0,1]),\psi(0)=\psi(1)\right\}
\label{eqn: Dom(p)}
\end{equation}
With this domain, $\hat p$ is symmetric, but it remains to be shown that it is (essentially) self-adjoint\footnote{{An operator is essentially self-adjoint if the corresponding closure operator $\hat p^{cl}$ is self-adjoint. See Hall \cite{hall2013quantum} for details.}}.}

{The coordinate transform $q\mapsto\theta=2\arcsin(\sqrt{q})$ in \cref{eqn:Transform} gives rise to a unitary mapping $U$ between the $L^2$ spaces in $q$ (on $[0,1]$) and $\theta$ (on $[0,\pi]$). Explicitly, we write,
\begin{equation*}
U:L^2([0,1],d\mu(q))\rightarrow L^2([0,\pi],d\theta), \quad (U\psi)(\theta)=\psi\left[\sin^2\left(\frac{\theta}{2}\right)\right]
\end{equation*}
The new Lebesgue measure $d\theta$ is equal to the weighted Lebesgue measure $d\mu(q)=\sqrt{|g|}dq$ given in the Riemannian volume form. In terms of $\partial_\theta$, the momentum operator becomes,
\begin{equation*}
\hat p =-i\hbar \sqrt{q(1-q)}\frac{\partial}{\partial q}=-i\hbar \sqrt{q(1-q)}\frac{d\theta}{dq}\frac{\partial}{\partial \theta}=-i\hbar\frac{\partial}{\partial \theta}.
\end{equation*}
Written this way, $\hat p$ has domain 
\begin{equation*}
\operatorname{Dom}(\hat p)=\left\{\psi\mid U\psi\in C^2([0,\pi]), (U\psi)(0)=(U\psi)(\pi)\right\}. 
\end{equation*}
It is straightforward to show that the differential operator $-i\hbar\frac{\partial}{\partial\theta}$ with such a domain is essentially self-adjoint, as it has an orthonormal eigenbasis (see Chapter 9 of \cite{hall2013quantum} for details).}

{Since $\hat p$ is essentially self-adjoint, the domain of the adjoint $\hat p^{*}$ must be the same as the domain of the closure $\hat p^{cl}$. The closure operator has graph equal to the closure in $L^2\times L^2$ of the graph of $\hat p$. Equivalently, the domain of $\hat p^{cl}$ is the closure in $L^2$ of $\operatorname{Dom}(\hat p)$ with respect to the graph norm generated by $\hat p$. Since $U$ is a unitary mapping, we can write, 
\begin{align*}
\norm{\psi}^2_{\text{graph}}&=\norm{\psi}^2_{L^2([0,1],d\mu(q))}+\norm{\hat p\psi}^2_{L^2([0,1],d\mu(q))}\\
&=\norm{U\psi}^2_{L^2([0,\pi],d\theta)}+\norm{\frac{d}{d\theta}(U\psi)}^2_{L^2([0,\pi],d\theta)}=\norm{U\psi}^2_{H^1([0,\pi])},
\end{align*}
where $H^1$ is the Sobolev space of square-integrable functions whose weak (i.e. distributional) first derivatives are also square-integrable. Here, we have set $\hbar=1$ briefly for convenience, as its value has no effect on the above derivation. Thus, the domain of $\hat p^{cl}$ can be thought of as either the completion of $\operatorname{Dom}(\hat p)$ with respect to $||\cdot||_{\text{graph}}$ or the completion of $U\left(\operatorname{Dom}(\hat p)\right)$ with respect to $\norm{\cdot}_{H^1}$. Since $\hat p$ is essentially self-adjoint, we have that the momentum adjoint operator $\hat p^*$ has domain,
\begin{equation}
\operatorname{Dom}(\hat p^*)=\operatorname{Dom}(\hat p^{cl})=\left\{\psi\in L^2([0,1],d\mu(q))\mid U\psi \in H^1([0,\pi]),\psi(0)=\psi(1)\right\},
\label{eqn: Dom(p*)}
\end{equation}
where the periodicity requirement is equivalently written as $(U\psi)(0)=(U\psi)(\pi)$. It is clear from this analysis that the momentum operator $\hat p$ with domain as defined in \cref{eqn: Dom(p)} is not self-adjoint. To make the momentum self-adjoint, we can simply define $\operatorname{Dom}(\hat p)$ to be the adjoint domain given by \cref{eqn: Dom(p*)}.}

{We now turn our attention to $\hat p^2$. Letting $\phi,\psi\in\operatorname{Dom}(\hat p)$, we have,
\begin{align*}
\langle \phi,\hat p^2\psi\rangle&=-\hbar^2\int_0^1\overline{\phi(q)}\sqrt{q(1-q)}\frac{\partial }{\partial q}\left(\sqrt{q(1-q)}\frac{\partial\psi }{\partial q}\right)\frac{dq}{\sqrt{q(1-q)}}\\
&=-\hbar^2 \left.\overline{\phi(q)}\sqrt{q(1-q)}\frac{\partial\psi }{\partial q}\right|_0^1+\hbar^2\int_0^1\frac{\partial\overline\phi}{\partial q}
\sqrt{q(1-q)}\frac{\partial\psi }{\partial q}dq.
\end{align*}
In order for the boundary term to vanish, we may set,
\begin{equation*}
\lim_{q\to 0}\sqrt{q(1-q)}\frac{\partial\psi }{\partial q} = \lim_{q\to 1}\sqrt{q(1-q)}\frac{\partial\psi }{\partial q}.
\end{equation*}
This is the same as requiring,
\begin{equation*}
    \frac{d(U\psi)}{d\theta}\bigg|_{\theta=0}=\frac{d(U\psi)}{d\theta}\bigg|_{\theta=\pi}.
\end{equation*}
Having thus restricted to functions whose transformed derivatives are periodic, we can write,
\begin{align*}
\langle \phi,\hat p^2\psi\rangle&=\hbar^2\int_0^1\frac{\partial\overline\phi}{\partial q}
\sqrt{q(1-q)}\frac{\partial\psi }{\partial q}dq.\\
&=\hbar^2 \left.\sqrt{q(1-q)}\frac{\partial\overline\phi}{\partial q}\psi(q)\right|_0^1+\int_0^1\overline{\left(-\hbar^2\frac{\partial}{\partial q}(\sqrt{q(1-q)}\frac{\partial\phi}{\partial q})\right)}\psi(q)dq\\
&=\int_0^1\overline{(\hat p^2\phi)}\psi(q)\frac{dq}{\sqrt{q(1-q)}}=\langle \hat p^2\phi,\psi\rangle.
\end{align*}
Thus, we see that $\hat p^2$ and by extension $\hat T$ are symmetric operators on this domain.}

{Let us briefly set $\tfrac{\hbar^2}{2m}=1$ for convenience. Then, under the transform in \cref{eqn:Transform}, $\hat p^2=\hat T=\frac{\partial^2}{\partial\theta^2}$. Written in this form, it is clear that these too are essentially self-adjoint operators, so $\operatorname{Dom}(\hat T^{cl})=\operatorname{Dom}(\hat T^*)$. To find this domain, we look for the closure under the graph norm generated by $\hat T$.
\begin{align*}
\norm{\psi}^2_{\text{graph}}&=\norm{\psi}^2_{L^2([0,1],d\mu(q))}+\norm{\hat T\psi}^2_{L^2([0,1],d\mu(q))}\\
&=\norm{U\psi}^2_{L^2([0,\pi],d\theta)}+\norm{\frac{d^2}{d\theta^2}(U\psi)}^2_{L^2([0,\pi],d\theta)}=\norm{U\psi}^2_{H^2([0,\pi])}-\norm{\frac{d}{d\theta}(U\psi)}^2_{L^2([0,\theta],d\theta)},
\end{align*}
where $H^2$ is the Sobolev space of square-integrable functions whose weak first and second derivatives are also square-integrable. Since the domain for $\hat p$ was already restricted to $H^1$ functions (after the transform), the domain of $\hat T^*$ is equivalently thought of as the completion of $\operatorname{Dom}(\hat T)$ under the graph norm or the completion of $U(\operatorname{Dom}(\hat T))$ under the $H^2$ norm. Explicitly, we have,
\begin{multline*}
\operatorname{Dom}(\hat T^*)=\left\{\psi\in L^2([0,1],d\mu(q))\mid U\psi \in H^2([0,\pi]),\right.\\
\left.(U\psi)(0)=(U\psi)(\pi), (U\psi)'(0)=(U\psi)'(\pi)\right\}.
\end{multline*}
Let $H^2_{\text{per}}([0,\pi])$ be the set of $H^2$ functions satisfying periodic boundary conditions. Since $H^2_{\text{per}}([0,\pi])$ is dense in $L^2([0,\pi])$, we have that $\operatorname{Dom}(\hat T)$ is dense in $L^2([0,1])$. Furthermore all operators defined thus far are defined on this domain. As such, we take it to be the space of test functions for our rigged Hilbert space.}

{It is worth noting that while periodic boundary conditions in the wavefunctions and their derivatives are required to make both momentum and energy self-adjoint, the kinetic energy can likewise be made self-adjoint if homogenous Dirichlet or Neumann boundary conditions are used instead, though momentum will no longer be an observable, as it fails to be self-adjoint. However, our information theoretic momentum does not necessarily correspond to a physical quantity. The lack of self-adjoint momentum may not be troubling so long as the energy, which can have a physical interpretation related to entropy, is still an observable.}




Finally, we now calculate the wavefunction for the free particle in Bernoulli space {with boundary conditions,}
{
\begin{equation}
\left\{
\begin{aligned}
&\psi(0) = \psi(1),\\
&\lim_{q\to 0}\sqrt{q(1-q)}\psi'(q) = \lim_{q\to 1}\sqrt{q(1-q)}\psi'(q).
\end{aligned}
\right.
\label{eqn:PeriodicBC}
\end{equation}
}
The Schr\"{o}dinger equation in this case is,
\begin{equation*}
    -\frac{\hbar^2}{2m}\Delta\psi = E\psi.
\end{equation*}
This yields,
\begin{equation*}
   {\psi(q)= c_1 \exp[2\kappa i\arcsin(\sqrt q)], \quad \kappa\equiv\frac{\sqrt{2mE}}{\hbar}.}
\end{equation*}
{Accounting for the boundary conditions gives us the same eigenbasis from \cref{eqn: period_basis}, namnely,}
\begin{equation*}
   {\Psi_n(q) = \frac{1}{\sqrt{\pi}}\exp\left[4ni\arcsin(\sqrt{q})\right]}, \quad n\in\mathbb{Z}.
\end{equation*}
{Thus the energy of the free particle is also discretized into energy levels of the form,
\begin{equation*}
E_n=\frac{2\hbar^2n^2}{m}, \ \ n \in\mathbb{Z}.
\end{equation*}}

{Were we to choose homogenous Dirichlet boundary conditions instead, we would get the expected eigenbasis $\Phi_n(q)$ as in \cref{eqn: dirichlet_basis}. This form of the} wavefunction for a free particle in {Bernoilli} space is analogous to the sinusoidal wavefunctions in the standard {Euclidean} \textit{particle in a box} \cite{griffiths2018introduction} infinite square well. The energy levels {are then} of the expected form,
\begin{equation*}
   E_n=\frac{\hbar^2n^2}{2m},\ \ n\in\mathbb{Z}.
\end{equation*}
{As noted above, homogenous Dirichlet boundary conditions only make the energy self-adjoint, but momentum will not be observable. In the following discussion of quantum harmonic oscillators, we will only be considering periodic boundary conditions.}

\section{Quantum Harmonic Oscillators}\label{sec:HarmonicOscillator}
The kinetic energy action in \cref{eqn:KEAction} yields the geodesic square distance,
\begin{equation*}
    {D_G(q,q') = \frac{1}{2}\left(2\arcsin(\sqrt{q}) - 2\arcsin(\sqrt{q'})\right)^2.}
\end{equation*}
The geodesic harmonic oscillator in this geometry is then given by the Lagrangian,
\begin{equation*}
    \mathcal{L}_G = \frac{m}{2}\frac{\dot{q}^2}{q(1-q)}-\frac{k}{2}D_G(q,q').
\end{equation*}
Deriving its Hamiltonian and quantizing it is then straightforward, and using the foregoing analysis, it is immediately clear {that in Bernoulli space we} will obtain a modification to the standard quantum harmonic oscillator {on a circle.} 

Alternatively, when studying non-linear spring dynamics in Bernoulli space, Goehle and Griffin \cite{goehle2024dynamics} use the Kullback-Liebler divergence as their square distance to obtain the classical Hamiltonian,
\begin{equation*}
    \mathcal{H} = \frac{1}{2m}q(1-q)p^2 + {\frac{k}{2}D_{KL}(q'\vert q)}.
\end{equation*}
This Hamiltonian and its corresponding Lagrangian were chosen to better represent a time-varying {distribution (or belief)} subject to external information (in the form of $q'$) \cite{goehle2024dynamics}. {In this case, the potential energy term represents the energy required to process new information, which pulls on the distribution (particle) like a spring.}

Before proceeding further, it is instructive to consider a dimensional analysis since the divergence is in nats as the square distance unit \cite{crooks2007measuring} and the kinetic energy action has units of nats per second. The structure of the equations implies that the kinetic energy (without mass) is in nats per second squared. To build clear and consistent physics in information space, we must define a new inertial unit unique to this space, which we whimsically call the `nert'. The spring constant $k$ then has units of nerts per second squared. Physics can then be conducted in this space in units of nats, nerts, and seconds, which effectively correspond to the three fundamental units in ordinary space: (square) meters, kilograms, and seconds.

\begin{figure}[htbp]
\centering
\includegraphics[width=0.8\textwidth]{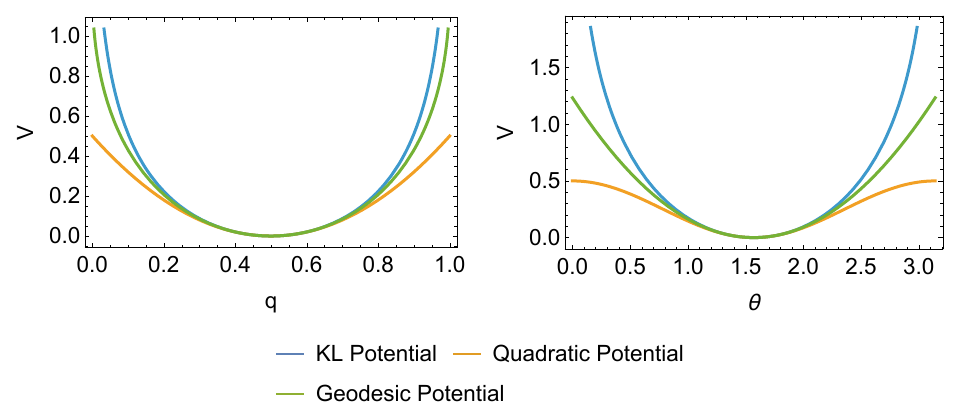}
\caption{{(Left) The KL, Geodesic, and quadratic potentials in Bernoulli space with $q' = \tfrac{1}{2}$. (Right) The same potentials in Euclidean space.}}
\label{fig:Potentials}
\end{figure}

The nonlinearity of the KL divergence unfortunately seems to render any closed-form analysis of the given Hamiltonian intractable. {Interestingly,} {at} second-order approximation, we have,
\begin{equation*}
    {D_G(q',q) \sim D_{KL}(q'\vert q)} \sim {2}(q' - q)^2,
\end{equation*}
which resembles the Euclidean square distance. {We} consider this quadratic potential {as a low-order approximation of the two information theoretic potentials}. {All three potentials are compared in Bernoulli space and Euclidean space in \cref{fig:Potentials} using $q' = \tfrac{1}{2}$.}

As before, we have $\hat{p} = -i\hbar\sqrt{q(1-q)}\frac{\partial}{\partial q}$. We recover the time-invariant Schr\"{o}dinger equation,
\begin{equation*}
    -\frac{\hbar^2}{2m}\sqrt{q(1-q)}\frac{d}{dq}\left[\sqrt{q(1-q)}\frac{d\psi}{dq}\right] + {k}(q'-q)^2\psi = E\psi,
\end{equation*}
where we require the {periodic} boundary conditions ({see \cref{eqn:PeriodicBC}}). In general, this does not have a simple solution for arbitrary $q'$. However, if we set $q' = \tfrac{1}{2}$, corresponding to the centre of the space, and apply the transform in \cref{eqn:Transform}, we obtain the transformed Schr\"{o}dinger equation,
\begin{equation*}
    -\frac{\hbar^2}{2m}\frac{d^2\psi}{d\theta^2} + \frac{k}{{4}}\cos^2(\theta)\psi = E\psi,
\end{equation*}
which reduces to the Mathieu differential equation,
\begin{equation*}
    \frac{d^2\psi}{d\theta^2} + \left[\frac{{8}mE - km}{{4}\hbar^2} - \frac{km}{{4}\hbar^2}\cos(2\theta)\right]\psi = 0.
\end{equation*}
We can relate this directly to the quantum pendulum \cite{aldrovandi1980quantum,baker2002quantum}, given by the Mathieu differential equation,
\begin{equation*}
    \frac{d^2\psi}{d\eta^2} + \left[\frac{2mEl^2-2m^2gl^3}{\hbar^2}-\frac{2m^2gl^3}{\hbar^2}\cos(\eta)\right]\psi = 0.
\end{equation*}
Letting $2\theta = \eta$, the equation becomes,
\begin{equation*}
    \frac{d^2\psi}{d\theta^2} + \left[\frac{8mEl^2-8m^2gl^3}{\hbar^2}-\frac{8m^2gl^3}{\hbar^2}\cos(2\theta)\right]\psi = 0.
\end{equation*}
Relating the two equations termwise implies,
\begin{equation*}
    \frac{km}{{4}\hbar^2}\leftrightarrow\frac{8m^2gl^3}{\hbar^2}\Longrightarrow k \leftrightarrow {32}mgl^3,
\end{equation*}
and
\begin{equation*}
    \frac{{8}mE - km}{{4}\hbar^2} \leftrightarrow \frac{8mEl^2-8m^2gl^3}{\hbar^2}.
\end{equation*}
Substituting in the value for $k$ and solving for $l$ gives,
\begin{equation*}
    l \leftrightarrow \frac{1}{2},
\end{equation*}
which is consistent with our assumption that $q'$ is centred in Bernoulli space and shows the direct transformation from a quantum oscillator using a na\"{i}ve Hooke's law potential in Bernoulli space to the quantum pendulum in Euclidean space. Interestingly, in this transformation, the units of $k$ are in $\mathrm{m}^4\mathrm{kg}/s^2$. That is, we see distance to the fourth power appearing in the units of $k$, which is why we use a relation sign $\leftrightarrow$ rather than an equals sign.  

\begin{figure}[htbp]
\centering
\includegraphics[width=1.0\textwidth]{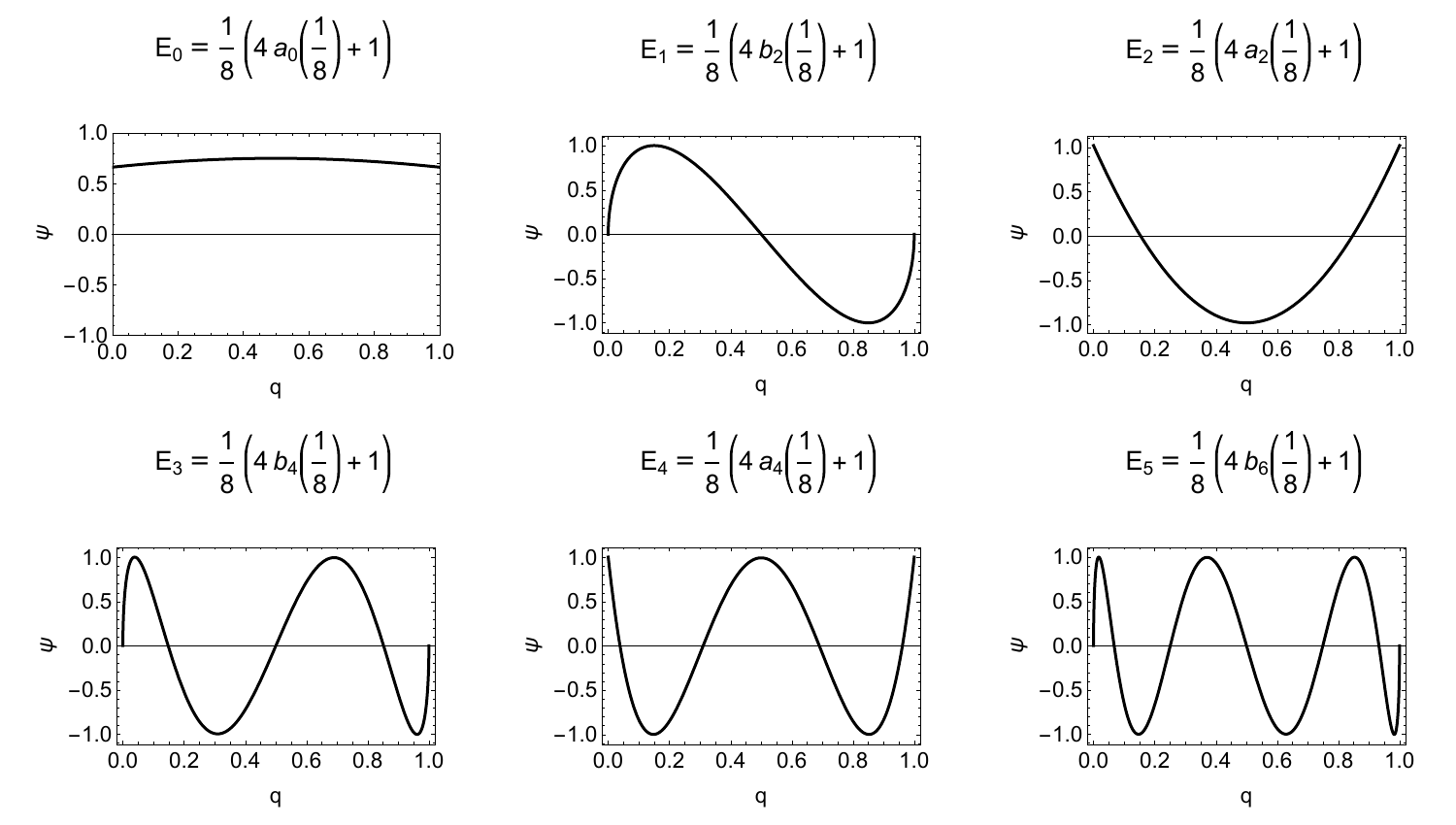}
\caption{{Harmonic oscillator energy levels assuming a quadratic square distance function in Bernoulli space. The wave function in each energy level corresponds either to a Mathieu $C$ or $S$ function.}}
\label{fig:OscillatorEnergyLevels}
\end{figure}

Solving our Mathieu differential equation gives us
\begin{equation*}
\psi(\theta) = c_1 C\left[\frac{{8}mE - km}{{4}\hbar^2},\frac{km}{8\hbar^2},\theta\right]+c_2
   S\left[\frac{{8}mE - km}{{4}\hbar^2},\frac{km}{{8}\hbar^2},\theta\right].
\end{equation*}
Here, $S$ and $C$ are the Mathieu functions of the first kind. {Imposing} {periodic} boundary conditions (as in the standard quantum pendulum \cite{aldrovandi1980quantum}), {requires the use of the Mathieu characteristic functions $a_n(\mathfrak{q})$ and $b_n(\mathfrak{q})$, with $n$ a non-negative integer, to recover values for the energy, where,
\begin{equation*}
    \mathfrak{q} = \frac{km}{8\hbar^2}.
\end{equation*}
Here we use $\mathfrak{q}$ in place of the usual $q$, because we are already using $q$ as a coordiante. We note for fixed $\mathfrak{q} > 0$ \cite{blanch1972handbook},
\begin{equation*}
    a_0(\mathfrak{q}) < b_1(\mathfrak{q}) < a_1(\mathfrak{q}) < \cdots
\end{equation*}
To enforce our required $\pi$-periodicity (as opposed to $2\pi$-periodicity), our eigenvalues are given by $a_{2n}(\mathfrak{q})$ and $b_{2n}(\mathfrak{q})$. Thus our energy levels and wavefunction basis has form,
\begin{gather*}
    E_0 = \frac{4\hbar^2a_0(\mathfrak{q})+km}{8m} \longleftrightarrow \psi_0(q) = C\left[\frac{8mE_0 - km}{4\hbar^2},\frac{km}{{8}\hbar^2},2\arcsin\left(\sqrt{q}\right)\right],\\
    E_1 = \frac{4\hbar^2b_2(\mathfrak{q})+km}{8m} \longleftrightarrow \psi_1(q) = S\left[\frac{8mE_1 - km}{4\hbar^2},\frac{km}{{8}\hbar^2},2\arcsin\left(\sqrt{q}\right)\right],\\
    E_2 = \frac{4\hbar^2a_2(\mathfrak{q})+km}{8m} \longleftrightarrow \psi_2(q) = C\left[\frac{8mE_2 - km}{4\hbar^2},\frac{km}{{8}\hbar^2},2\arcsin\left(\sqrt{q}\right)\right],\\
    \vdots
\end{gather*}
Notice that the energy term $E_n$ is a function of $a_n$ for even $n$ and $b_{n+1}$ for odd $n$. The first few stationary states are shown in \cref{fig:OscillatorEnergyLevels} for $\hbar = m = k = 1$.}

\begin{figure}[htbp]
\centering
\includegraphics[width=0.45\textwidth]{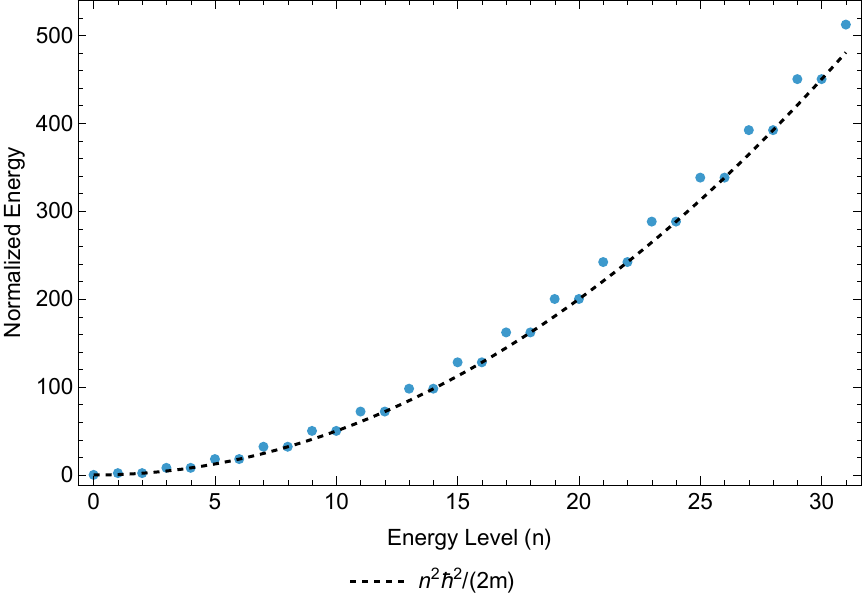}
\caption{{The energy levels of the Bernoulli quantum harmonic oscillator with quadratic potential exhibit a paired energy behaviour and grow quadratically.}}
\label{fig:OscillatorEnergyLevelsValue}
\end{figure}

{A plot of the energy levels is shown in \cref{fig:OscillatorEnergyLevelsValue}.}
{Notice all energy levels but the ground state essentially come in pairs corresponding to the (slightly confusing) $b_{n}(\mathfrak{q})$ and $a_{n+1}(\mathfrak{q})$ for $n \geq 1$. For large $n$, the characteristics values obey $a_n(\mathfrak{q}), b_{n}(\mathfrak{q}) \sim n^2$ \cite{blanch1972handbook}. Consequently, it follows that, 
\begin{equation*}
     E_n \sim \frac{\hbar^2n^2}{2m},
\end{equation*}
for large $n$. This is interestingly consistent with the energy of a particle in a box, as discussed.} 

\begin{figure}[htbp]
\centering
\includegraphics[width=0.8\textwidth]{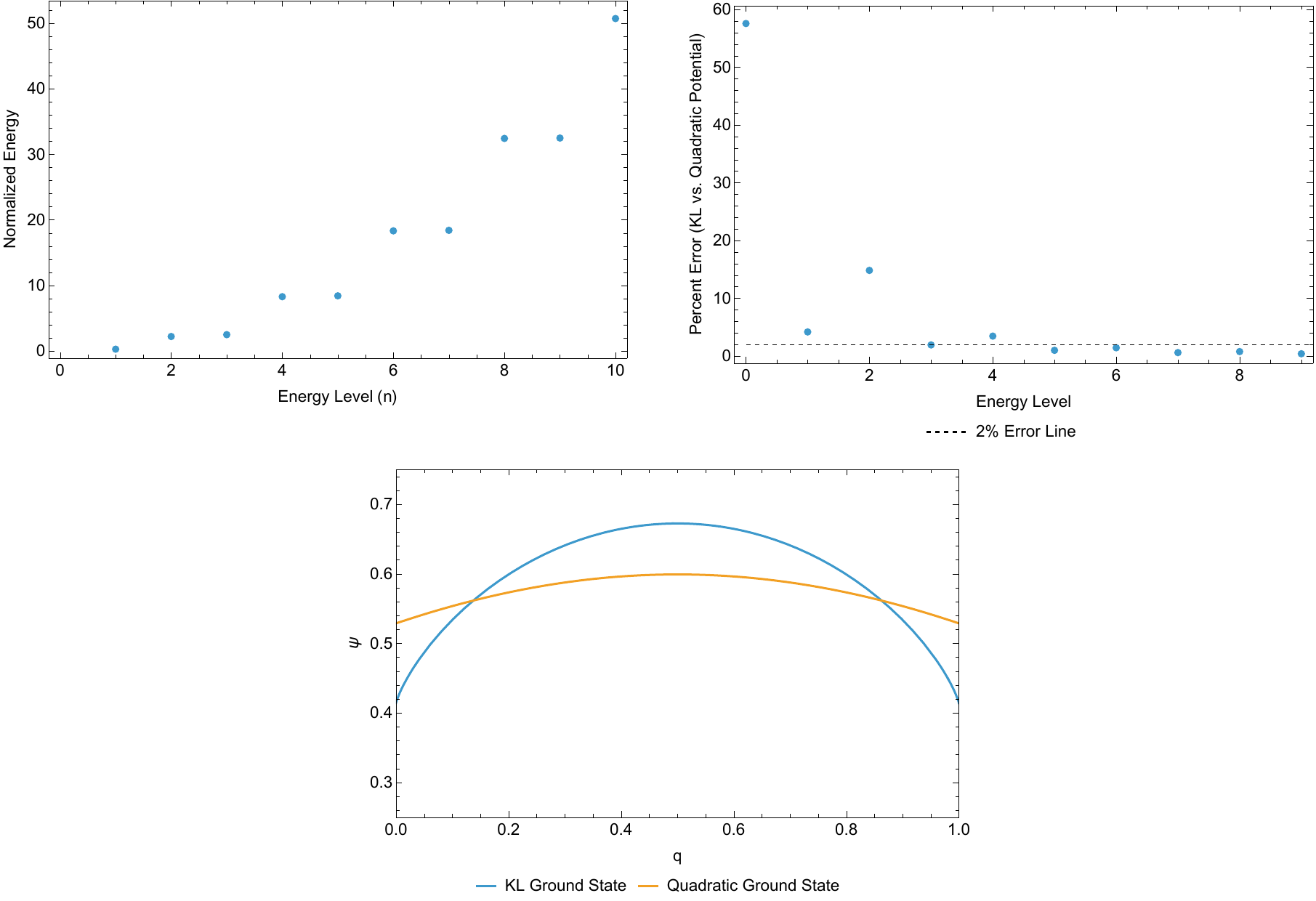}
\caption{{(Top Left) The first ten energy levels of the KL potential harmonic oscillator. Note these energy levels also come in pairs. (Top Right) The approximation error is low as $n$ becomes large. Lower energy levels have larger approximation errors. (Bottom) The ground state wavefunction in the quadratic potential and the KL potential are compared.}}
\label{fig:KLEnergy}
\end{figure}

{\cref{fig:Potentials} make it clear that the energy levels of the quadratic approximation are lower bounds for the general information theoretic potentials (square distances). We show numerically that (except for the ground state), the energy levels of the quadratic potential provide a good approximation for those of the KL potential in \cref{fig:KLEnergy}.}
{Note that the KL divergence potential also generates pairs of energy levels with similar energy, thus making them qualitatively similar to the energy levels of the quadratic potential function. This is illustrated in  \cref{fig:KLEnergy} (Top Left). For $n>4$, the energy levels of the quadratic potential approximate the energy levels of the KL potential, which are shown in \cref{fig:KLEnergy} (Top Right). The greatest difference occurs in the ground state, seen in \cref{fig:KLEnergy} (Bottom). Thus, the quadratic potential in Bernoulli space provides an approximation for the general KL or geodesic potentials that is similar to a perturbative analysis. Unfortunately, if $q' \neq \tfrac{1}{2}$, there is no direct transformation to a quantum pendulum. Instead, the resulting system requires numerical analysis or could be analyzed perturbatively by setting $q' = \tfrac{1}{2} + \epsilon$ and expanding. However, this does not provide any additional insight as we will simply obtain perturbed Mathieu functions.
}

{While there is currently no direct physical interpretation for the quantum dynamics in Bernoulli space, this work is related to superstatistics. In superstatistics, the parameters of a distribution (e.g., the inverse temperature in the Boltzmann distribution) are allowed to vary according to their own distributions \cite{beck2003superstatistics,cohen2004superstatistics}. Quantum superstatistics assumes these variations are quantum mechanical in nature \cite{ourabah2017entanglement,ourabah2017quantumA,hassanabadi2021superstatistics}. Such models can be used to describe out-of-equilibrium thermodynamic systems with appropriate quantum extensions. In particular, the model we study is equivalent to a two state Boltzmann distribution with a quantum mechanically varying parameter. 
}

\section{Conclusions and Future Directions}\label{sec:Conclusion}
In this paper, we have studied both classical and quantum physics on an information theoretic manifold arising from a Bernoulli distribution, extending work from Goehle and Griffin \cite{goehle2024dynamics}. Since information manifolds admit not only natural square distance geodesics, but also consistent divergences, we showed that there are many ways to define a natural quantum harmonic oscillator on these manifolds and that the second-order approximation{s} of {both} the Kullback-Leibler potential {and the geodesic potential} (the na\"{i}ve quantum harmonic oscillator in Bernoulli space) produce an oscillator equivalent to the quantum pendulum in Euclidean space. This analysis was enabled by a quantization of the momentum operator identified in prior work \cite{goehle2024dynamics}.

There are several future directions to consider. Determining whether these results may be of use in an experimental context is obviously of interest. Extensions of this work to more general information spaces could yield additional insights into quantum mechanics in highly curved spaces, since all information manifolds seem to exhibit curvature {(of some kind)} corresponding to areas of low variance. Finally, a more in-depth investigation into the relationship between this work and {other} more general work in quantum geometry \cite{facchi2010classical} {or thermodynamic length \cite{crooks2007measuring}}  may yield further insights.

\section*{Acknowledgments}
S.G. was supported in part by the Schreyer Honors College at the Pennsylvania State University. C.G. was supported in part by the Office of Naval Research under Naval Sea Systems agreement number N0002425F8306.

\section*{Data Availability Statement}
The data that support the findings of this study are available within the article [and its supplementary material].

\section*{Statement on the use of Artificial Intelligence}
This paper was written entirely by the authors without the use of artificial intelligence. Claude was used to generate information on quantum superstatistics, which was then verified in a separate literature review. 

\bibliographystyle{iopart-num}
\bibliography{main}

\end{document}